# Toward a Geopolitical Crisis Observatory:
# Diagnosing Systemic Risk in News Flows Using Complex Systems Science

**D. Sornette**
Chair Professor and Dean of the Institute of Risk Analysis, Prediction and Management (Risks-X)
Southern University of Science and Technology (SUSTech), Shenzhen, 518055, China
Professor em. ETH Zurich and UCLA

**Abstract**: Complex-systems science provides media institutions with a rigorous framework to move from reactive reporting to **anticipatory diagnosis**. Critical events are understood as regime shifts emerging from the interplay between **endogenous dynamics and exogenous shocks**. Detecting such transitions requires identifying structured precursors, such as changes in correlations, amplification, persistence, and endogeneity, rather than relying on raw signal intensity. Recognizing **dragon-king events** as regime-generated outliers and incorporating **non-normal transient amplification** are essential, as is accounting for **organizational concealment of risk**. A geopolitical crisis observatory would diagnose when systems enter states of heightened susceptibility to cascading disruptions. While state actors are already developing such observatories for strategic purposes, media institutions remain largely reactive. This gap creates a strategic opportunity: leveraging open data and AI embedded within the complex-systems framework developed above, media organizations could transform journalism from reporting to diagnosis, delivering early-warning indicators, scenario-based risk maps, and transparent, data-driven narratives within a new **Geopolitical Risk Intelligence Platform**.

**Empirical validation: from concept to operational reality**

A recent real-world case illustrates that the core principles of a Geopolitical Crisis Observatory are not hypothetical but already being implemented in strategic contexts. As reported in April 2026, Chinese actors are leveraging artificial intelligence applied to open-source data (OSINT) to monitor, reconstruct, and anticipate U.S. military operations in Iran in quasi real time. This example demonstrates that the fusion of multi-modal open data (satellite imagery, maritime tracking, social media, logistics signals) with machine learning allows not only retrospective analysis but forward-looking inference of operational intentions. In other words, what is proposed below for media institutions already exists in a more advanced form in military-strategic contexts. The key challenge is therefore not feasibility but institutional translation: how to adapt such capabilities into a civilian, journalistic, and public-interest observatory.

**1. The problem is not forecasting "events" in the journalistic sense, but diagnosing systems that are becoming susceptible to regime shifts**

The first task is conceptual clarification. The expression "critical event" is too imprecise to support a serious observatory unless one first distinguishes between an event as a newsworthy occurrence and a critical transition as understood in the science of complex systems. In my research, the central object is not the isolated event but the **dynamical regime** that precedes and conditions it. A crisis, a crash, a war escalation, a wave of social unrest, or a sudden breakdown of institutional legitimacy should not be treated merely as an ex post "surprise." In many important cases, such outcomes are the visible culmination of a longer endogenous buildup of stress, imitation, positive feedbacks, delayed adaptation, hidden fragilities, and information distortions. This is the logic developed in my work on financial crashes, dragon-kings, catastrophic events, and man-made catastrophes. The important question is therefore not "Can one predict the exact date of the next geopolitical crisis?" but rather "Can one diagnose that a socio-economical-political system has entered a phase in which it has become hypersensitive to perturbations and therefore liable to produce a burst of activity and/or a critical transition?"

That distinction is decisive for media institutions. If journalism limits itself to identifying exogenous triggers, it remains hostage to the superficial chronology of events. But if media institutions build the capacity to monitor endogenous stress accumulation, changing couplings among narratives, emotional polarization, organizational concealment of risk, and the transient amplification properties of the information ecosystem itself, then they can move from reactive coverage to anticipatory diagnosis. This would be analogous, mutatis mutandis, to the Financial Crisis Observatory that I founded in 2008, whose purpose was the real-time diagnosis of unsustainable bubble regimes in financial markets. The FCO explicitly frames its mission as testing whether bubble phases exhibit measurable signatures of instability and potential predictability (see https://sornette.finance for the present operational professional version). The same philosophy can be extended to the social and geopolitical sphere, provided the objects being monitored are defined correctly.

## 2. The relevant distinction is endogenous, exogenous, and mixed crises

A rigorous media-oriented early-warning framework must start from the endogenous-exogenous decomposition. This is one of the most useful conceptual and empirical distinctions I have developed with collaborators across several domains. In the endo-exo framework on social response functions, endogenous bursts and exogenous shocks were shown to leave different relaxation signatures in collective attention data. That result is not an anecdotal curiosity; it provides a prototype for how one can infer the causal structure of observed bursts from their temporal shape. Likewise, in (Filimonov and Sornette, 2012; 2015; Bicchetti et al., 2014), high endogeneity in financial and commodity markets was quantified using self-exciting point-process approaches, showing that what looks like "news-driven" volatility is often in large part generated by the system's internal reflexive dynamics.

Applied to media-based analysis of geopolitical crises, this means that not all crises should be treated in the same way. Some are predominantly exogenous. A military strike, a sudden assassination, an earthquake, or an unambiguously external policy shock may trigger massive information cascades and destabilisation, but the trigger is genuinely outside the internal news-attention dynamics. In this case, the degree of criticality of the underlying system is not revealed by the trigger itself but by the characteristics of the response: its amplitude, its persistence, its relaxation pattern and possibly the occurrence of regime shifts. A highly susceptible system will exhibit disproportionately large and long-lasting reactions, with slow decay and strong secondary reverberations, reflecting the activation of internal feedback loops. This leads naturally to the notion of an "exo-critical" situation, in which the system's endogenous state amplifies and sustains the impact of external shocks, transforming a simple perturbation into a potential regime-shifting event.

Other crises are predominantly endogenous. Financial bubbles, bank runs, organizational implosions, legitimacy collapses, rumors that become self-fulfilling, urban unrest incubated by long-standing grievances, power competition or ethnic minority frustrations are generated largely through internal positive feedback loops. Between these two lies the most important class: mixed crises, in which a long endogenous maturation creates a metastable system and a small external perturbation precipitates the transition. In such cases, the perturbation is not the true cause in any deep dynamical sense. It is only the last drop in a system already made critical by its internal organization.

This is precisely where many media analyses fail. They over-attribute causality to the visible trigger and underweight the long endogenous preparation. In the lead-up to many crises, one observes superposition of stressors: debt accumulation, erosion of institutional trust, environmental degradation, persistent asymmetries in information access, strategic concealment, and narrative simplification. A media observatory worthy of the name should therefore aim to measure not only trigger-events but also the evolving **susceptibility function** of the system. In other words, one wants to estimate when the same perturbation would produce only a limited fluctuation and when, by contrast, it would cascade into a full regime shift.

### Illustration: the Iran conflict as a live endo-exo laboratory

The Iran theater provides a striking illustration of the endogenous-exogenous decomposition. While military strikes constitute exogenous triggers, the Chinese approach mentioned above does not focus on these triggers per se, but on the evolving configuration of signals that precede them. By aggregating logistical indicators (fuel movements, aircraft clustering), satellite observations, and open-source communications, AI systems identify structured precursor patterns that correspond exactly to what complex-systems science defines as endogenous maturation. This confirms that what appears as an exogenous "event" in media narratives can be partially reconstructed as the culmination of measurable endogenous preparation.

## 3. What the Financial Crisis Observatory has really contributed was a methodology for diagnosing unsustainable regimes, not a market metaphor

The relevance of the Financial Crisis Observatory for media is often misunderstood if reduced to a loose analogy such as "finance has bubbles, politics has crises." The deeper lesson is methodological. In financial markets, the LPPLS framework that I have developed formalized the idea that bubbles are not defined by high prices alone but by a specific dynamical regime: faster-than-exponential growth generated by positive feedbacks among agents, decorated by accelerating oscillations due to competition between trend followers and value investors across hierarchical timescales (Sornette, 2002; Sornette and Cauwels, 2015). This allows us to routinely diagnose a loss of dynamical sustainability and to constrain windows of likely regime change. The FCO institutionalized this diagnostic philosophy at scale (Sornette et al., 2009; 2010; Woodard et al., 2010). (see https://sornette.finance for the present operational professional version).

For a media-based observatory, the direct analogue is not "price" but a vector of collective-state variables: attention intensity, emotional valence, semantic concentration, network clustering, synchronization across channels, opinion polarisation, background crime level and disruptions, and the degree of endogenous self-excitation. The scientific question becomes whether these observables display signatures analogous to bubble dynamics: accelerating growth in attention to a conflict frame; shortening oscillatory cycles between alarm and reassurance; rising cross-platform synchronization; intensifying emotional variance; and a progressive concentration of narratives around mutually reinforcing geopolitical themes. One should not naively transfer the LPPLS formula from asset prices to raw social-media and news volume. But one can transfer the logic of the method: identify **unsustainable positive-feedback regimes** and detect the approach to a critical zone before visible rupture. Recent developments in AI-enabled OSINT pipelines demonstrate that the equivalent of such diagnostics is already emerging in practice. In the Chinese case, large-scale ingestion of heterogeneous open data streams is used to reconstruct operational states and detect precursors of action. The methodological convergence is striking: both approaches rely on identifying structured deviations from baseline dynamics rather than reacting to isolated events. The difference is that AI systems now enable this at scale, in real time, and across multiple domains simultaneously.

A concrete example would be the escalation dynamics around a strategic chokepoint such as the Strait of Hormuz. If one wishes to build a serious observatory, one should not merely count mentions of "Hormuz," "Iran," or "oil prices." One should construct multiple synchronized time series: the volume and acceleration of conflict-related narratives; the ratio of fear to anger in Arabic-, English-, Persian-, and Hebrew-language discourse; the centralization of retweet or repost networks around military, state, or quasi-state accounts; the divergence between official language and crowd language; the appearance of lexical markers of inevitability, retaliation, humiliation, deterrence failure, and emergency logistics. One should also measure whether these streams are mainly responding to discrete exogenous announcements or whether they exhibit autonomous persistence after the triggering news has faded. The latter would indicate endogenization of the crisis dynamics inside the information system itself.

Prediction markets provide an obvious vehicle in which predictions of geopolitical events are transformed in speculative trading of financial contracts such as during the recent market craze on whether Iran's supreme leader, Ayatollah Ali Khamenei, would be out of power by March. As the U.S. moved soldiers and military equipment to bases in the Middle East, the prediction markets Kalshi and Polymarket set up the scenario as a "contract" for users to purchase. Payouts fluctuated based on the probability of Khamenei's ouster, and it grew into a $54 million market on Kalshi. More than in financial market, insider trading by actors "who are in the know" (such as in the US situation room) raise deep ethical issues but also the fascinating efficiency of financial markets to reflect information before it is revealed. My work in new design of prediction markets shows that such vehicles provide complementary means for diagnosing emerging crises (Sornette et al., 2020; Andraszewicz et al, 2020).

## 4. Weak signals are not just small signals; they are structured departures from baseline dynamics

Weak signals are often described in a vague managerial language that confuses faintness with relevance. Scientifically, a weak signal should be defined as a low-amplitude but dynamically structured departure from baseline that is statistically and mechanistically consistent with an evolving instability. The issue is not whether the signal is dramatic, but whether it belongs to a coherent precursor pattern.

The Chinese use of OSINT provides a concrete operational definition of weak signals. Individual data streams, such as ship transponders, satellite images, or social media posts, are indeed weak and fragmentary when taken in isolation. However, when aggregated and processed through machine learning, they reveal coherent patterns of military preparation. The signal is not in the magnitude of any single data stream, but in their structured correlation and synchronization. This corresponds exactly to the scientific definition proposed here: weak signals are structured deviations, not small amplitudes.

In my work on critical phenomena and catastrophic transitions, the relevant precursors include accelerations, increasing correlations, clustering, and changes in fluctuation structure. In social systems, the endo-exo framework that I have developed suggests that the shape of the response function itself can distinguish endogenous and exogenous bursts. In self-exciting systems, the branching ratio or effective endogeneity parameter provides an interpretable measure of how much activity is internally reproduced by previous activity. In a media observatory, weak signals should therefore be sought not only in levels but in **changes of dynamical organization**.

Consider a socio-political controversy whose total message volume remains moderate. A superficial newsroom would conclude that nothing exceptional is occurring. A crisis observatory, however, would examine whether the same moderate volume now exhibits longer memory, stronger clustering, slower decay after shocks, sharper synchronization across geographically separated communities, and increasing coupling between fear, anger, and conspiracy-related lexicons. That combination can signal a deep structural change even before headline volume explodes. Similarly, what matters is not just the presence of "anger" as an emotion, but whether anger becomes coupled with narratives of betrayal, urgency, and target-identification in a network topology conducive to rapid propagation.

This is where large-language-model tools and modern NLP can be useful, but only as measurement devices integrated into a dynamical framework. A scientifically serious observatory should extract time series for multiple emotional coordinates, not only the six canonical categories but broader affective and cognitive states such as anxiety, outrage, disgust, anticipation, uncertainty, confidence collapse, and calls for irreversible action. One should then study their lag structure, mutual information, and cross-coupling with actor mentions, policy frames, commodity prices, mobility constraints, military indicators, or environmental stress markers.

## 5. Operationalizing the Endogenous-Exogenous Framework: Self-Excitation, Reflexivity, and Event-Driven Dynamical Centrality Ranking

Moving from conceptual distinctions to implementation requires a class of models capable of disentangling external driving forces from internally generated dynamics. A natural and powerful framework is provided by response-function approaches and self-exciting processes, which allow one to reconstruct how past activity propagates and amplifies within a system. In earlier work on social response (Sornette and Helmstetter, 2003; Sornette et al., 2004; Sornette, 2005; Crane and Sornette, 2008), it was shown that the temporal relaxation following bursts of activity encodes information about their origin: externally triggered events exhibit different decay patterns than those sustained endogenously. This insight was later formalized in the context of financial markets, where self-exciting models were used to quantify reflexivity by estimating the fraction of activity that is generated by the system itself rather than by external news.

The same logic applies directly to geopolitical ecosystems probed view news, provided one generalizes to multivariate and heterogeneous settings. Instead of a single stream of events, one must consider a network of interacting streams corresponding to different emotional channels, linguistic communities, narrative frames, media platforms, or institutional sources. The key idea is to decompose observed activity into two components: an exogenous contribution reflecting external inputs, and an endogenous component arising from the propagation and amplification of past events across the system. The latter captures the recursive nature of attention dynamics, whereby information does not simply react to the world but actively reshapes its own future evolution.

Within this framework, the central object is no longer just the intensity of activity, but the structure of interactions governing how signals propagate. One can infer whether the system is becoming more self-excited, whether cross-channel contagion is strengthening, whether the memory of past events is lengthening, and whether the system is approaching a regime in which internal dynamics dominate over external driving. These properties provide direct diagnostics of systemic susceptibility: a highly endogenous system is one in which small perturbations can be repeatedly amplified, leading to large-scale cascades.

Recent developments further extend this approach by introducing the notion of **event-driven centrality dynamic ranking**, as formalized in the HawkesRank framework (Sornette et al., 2026). Rather than ranking actors or nodes based on static network topology or aggregate activity, HawkesRank evaluates their importance dynamically, based on their causal influence in triggering subsequent events. In other words, it identifies which sources, narratives, or actors are not merely visible, but are actively shaping the future evolution of the system through their propagation impact. This is particularly important in geopolitical contexts, where influence is inherently temporal and event-driven rather than represented by static networks.

Integrating such a measure into a geopolitical crisis observatory allows one to go beyond aggregate diagnostics and identify the loci of amplification within the system. For instance, one can detect when a particular narrative cluster, media outlet, or group of accounts becomes disproportionately effective at generating cascades, or when emotional signals such as fear or outrage acquire increased propagation power relative to other channels. One can also track shifts in influence across communities and languages, revealing how localized dynamics couple into global cascades.

The combination of endogenous-exogenous decomposition and event-driven centrality yields a much richer diagnostic language than conventional media analytics. Instead of stating that attention to a geopolitical issue is increasing, one can characterize whether the observed growth is externally driven or internally sustained, whether it is concentrated or distributed across actors, and whether the system is reorganizing toward a high-reflexivity regime. One can identify transitions in which a narrative moves from being a response to external events to becoming a self-sustaining process, maintained by internal feedback loops. One can also detect the strengthening of couplings between domains, such as the progressive entanglement of a geopolitical conflict with economic concerns like energy supply or insurance risk.

Such diagnostics are precisely what is needed for a scientifically grounded geopolitical crisis observatory. They transform raw data into interpretable measures of systemic dynamics, allowing one to assess not only what is happening, but how and why it is unfolding. This shift from descriptive monitoring to causal and dynamical inference is essential if media institutions are to develop genuine anticipatory capabilities rather than simply reacting to events after they occur.

The Chinese case further suggests that the next generation of observatories will not simply apply statistical models to pre-processed data, but will operate as continuous AI-driven inference systems. These systems integrate real-time ingestion of multi-modal data (satellite, logistics, social media), automated pattern recognition, predictive modeling of future states, and rapid updating of situational awareness. In this sense, a Geopolitical Crisis Observatory should be conceived as a civilian analogue of such systems, but governed by transparency, accountability, and editorial judgment rather than strategic secrecy.

**6. Dragon-kings matter because critical events are often not mere tail realizations but regime-generated outliers**

Another concept that should be central to a geopolitical observatory is the dragon-king. In (Sornette, 2009), I argued that many extreme events are not simply larger draws from the same distribution as ordinary events. They are generated by distinct amplification mechanisms and are therefore both exceptional and, in principle, more diagnosable than a pure "black swan" view would suggest. The key point for media is that a true crisis observatory must identify when a system is entering a domain in which dragon-kings become possible (Sornette and Ouillon, 2012).

In social and geopolitical systems, dragon-kings can correspond to sudden legitimacy collapses, abrupt cascades of violence, extraordinary surges in migration pressure, extreme market responses to political events, or information avalanches in which one narrative suddenly dominates all others. The observatory should not confuse ordinary volatility with a change in generative mechanism. The purpose is to detect when the system is no longer fluctuating around a stable attractor but is reorganizing toward a singular event generated by positive feedback, synchronization, and threshold effects.

This is also why simple trend extrapolation is insufficient. One does not want to predict tomorrow by extending yesterday. One wants to detect that the mechanism itself has changed. In some situations, the most informative observable will be the emergence of superlinear response (Sornette and Cauwels, 2015): each new perturbation produces a disproportionate increase in attention, outrage, or coordination. In others, it will be the appearance of log-periodic alternation between escalation and temporary calming, with shrinking intervals between oscillations (Johansen and Sornette, 1999; Sornette, 2002). In still others, it will be the collapse of diversity in narratives and the concentration of discourse into a few dominant antagonistic frames.

**7. Non-normal dynamics adds a new layer: collapse can occur through transient amplification before classical instability is visible**

My more recent work on non-normal phase transitions is especially promising for the construction of a next-generation geopolitical crisis observatory (Troude and Sornette, 2025; 2026; Troude et al.,2026). The central idea is that a system can remain asymptotically stable in the classical eigenvalue sense and still exhibit very large transient growth because of the geometry of its modes. In non-normal systems, eigenvectors are non-orthogonal, so perturbations can project onto combinations of modes that amplify strongly before any eventual decay. When this transient growth is large enough in the presence of nonlinearities or noise, it can induce regime change even though the conventional stability criterion has not yet been violated. This is a new universality class of phase transitions in which geometry, not only eigenvalue sign changes, governs the onset of qualitative change.

This is highly relevant to media and geopolitical systems because many geopolitical crises do not wait for a slow drift through a classical bifurcation. They may occur because the architecture of couplings among state actors, platforms, institutions, populations, national and international organisations, ethnic groups becomes such that certain perturbations are transiently but massively amplified. In operational terms, a crisis observatory should therefore monitor not only endogeneity and precursor acceleration but also the **pseudospectral vulnerability** of the information ecosystem.

What would that mean concretely? Suppose one represents a geopolitical ecosystem in a relatively stable status quo, represented mathematically by a linearized propagator around a moving baseline state. The nodes may correspond to power blocs, ethnic groups and/or nations. Even if the dominant measure of local stability remains below the classical instability threshold, the resolvent norm or related non-normal amplification measures can become large due to the presence of asymmetric and hierarchical interactions between actors. Then a moderate perturbation, say a leaked document, a bombing, a contradictory official statement, can align with an unstable transient direction in the dynamical space of the system and produce a system-wide surge in mobilization or over-reaction by some actors. The practical implication is profound: one can diagnose vulnerability not only from persistent buildup but from the changing geometry of couplings. This moves the observatory beyond simple instability thinking.

**8. Concealment of risk information and organizational pathology must be endogenous state variables of the observatory**

A social crisis observatory should not treat institutions merely as data sources. Institutions themselves are dynamical systems that can become crisis-generating through concealment, fragmentation, and suppression of contrary information. This is one of the central lessons of my book *Man-made Catastrophes and Risk Information Concealment* (Chernov and Sornette, 2016) and its sequels (Chernov and Sornette, 2020; Chernov et al., 2023), which analyzes how disasters are incubated by organizational pathologies long before the visible event. Concealment is not just lying. It includes complexity-induced opacity, communication breakdown, misaligned incentives, strategic silencing, and the filtering-out of inconvenient signals.

This has direct implications for geopolitical instabilities. A journalistically oriented observatory should monitor not only publicly available discourse but discrepancies between institutional narratives and peripheral signals. For instance, one can track divergence between official reassurances and field-level testimonies, between aggregate indicators and localized anomalies, or between declared preparedness and evidence of logistical strain. An observatory that ignores such discrepancies will systematically underestimate endogenous fragility and overestimate geopolitical stability. Indeed, one of the most robust findings across crises is that organizations often "know" more than they officially acknowledge, but their internal structure prevents weak signals from becoming actionable knowledge.

Thus, a serious observatory must incorporate organizational diagnostics. Which institutions show narrative rigidity? Which suppress variance in reporting? Which exhibit systematic delays between peripheral detection and central acknowledgment? Which repeatedly reclassify warnings as isolated incidents? These are not secondary sociological details; they are endogenous variables that help determine whether a society is approaching a crisis.

**9. From Military Intelligence to Media Intelligence: A Strategic Opportunity for Global Media Companies**

The phrase "geopolitical crisis observatory" can easily devolve into a fashionable but empty dashboard project. That would be a mistake. A genuine observatory should be built as a **hierarchical inference system** combining theory, data assimilation, and scenario diagnosis.

At the first layer, it would ingest heterogeneous time series: social-media content, mainstream media, geolocated attention data, market indicators, mobility proxies, environmental stress measures, and institutional communications. At the second layer, it would extract dynamical observables: endogeneity ratios, relaxation exponents, emotional factor dynamics, semantic polarization metrics, coupling matrices, cross-scale synchronization, and measures of non-normal amplification. At the third layer, it would estimate latent states: degree of systemic susceptibility, proximity to a critical transition, dominant mechanism of propagation, and whether the system is still in a linear-response regime or has entered a nonlinear amplification phase. At the fourth layer, it would produce not point predictions but conditional risk maps: if an exogenous shock of type X occurs, which latent instability channels are most likely to activate.

This is the right level of ambition. It is both scientifically grounded and institutionally useful. It also matches the spirit of the Financial Crisis Observatory (Sornette et al., 2009; 2010; Woodard et al., 2010; see https://sornette.finance for the present operational professional version), whose essential contribution was to formalize rigorous, real-time diagnostics of unstable financial market regimes rather than to generate simplistic predictions.

Recent developments in AI-driven exploitation of open-source data confirm that such an architecture is not hypothetical. In the context of the Iran conflict, large-scale systems already ingest satellite imagery, logistics signals, and social-media content to reconstruct operational states and anticipate future actions. The key novelty lies in the continuous integration of heterogeneous data streams into a dynamically updated inference pipeline. This demonstrates that the transition from observation to anticipation is technically feasible at scale.

This evolution creates a structural asymmetry: while state actors are already developing such observatories for strategic purposes, media institutions remain largely confined to reactive reporting. This gap creates a unique strategic opportunity for a large global media organization (e.g., Al Jazeera, BBC, Bloomberg, Reuters) to develop a **Geopolitical Crisis Observatory as a core editorial capability**. Such an observatory would:

(i**) Transform journalism from reporting to diagnosis**: Instead of covering events after they occur, the media would identify zones of systemic fragility and escalating risk.

**(ii) Leverage open data as a competitive advantage**: As demonstrated, high-value intelligence can be extracted from publicly available data when processed at scale.

**(iii) Provide differentiated, high-value content**: (i) early-warning dashboards for subscribers, (ii) scenario-based geopolitical risk maps and (iii)explanatory narratives grounded in data.

**(iv) Build trust through transparency**: Unlike military systems, a media observatory can expose its methodology, creating epistemic credibility.

**(v) Create a new product category**: A "Geopolitical Risk Intelligence Platform" combining journalism, data science, and complex systems analysis.

In this perspective, the Geopolitical Crisis Observatory is not only a scientific framework but also a **strategic institutional innovation**. It extends the logic of the Financial Crisis Observatory into the geopolitical domain while adapting it to a world in which open data and AI have fundamentally transformed the conditions of information production, competition, and anticipatory insight.

## 10. Case Study: The Iran–Hormuz System. From Exogenous Shock to Endogenous Global Amplification

The on-going 2026 Hormuz crisis provides a paradigmatic example of a mixed endo-exo system in which a geopolitical trigger interacted with a highly fragile global structure. The initial proximal phase was clearly exogenous: coordinated military strikes and subsequent escalation triggered the disruption of maritime flows through the Strait of Hormuz. However, from the perspective of a Geopolitical Crisis Observatory, the key question is not the occurrence of the strike but the **state of the system prior to it**.

Even before escalation, several structural fragilities were present. Approximately 20% of global oil supply transited through the Strait, creating a single-point-of-failure in the global energy system. Spare production capacity was limited, alternative transport routes were marginal, and supply chains had evolved toward "just-in-time" configurations with minimal buffers . In parallel, geopolitical tensions had already manifested through rising insurance premiums, increased rhetorical polarization, and growing strategic signaling around control of the Strait .

A properly configured observatory would have identified a progressive shift in news-flow dynamics. First, an increase in cross-domain coupling between military narratives and energy-market narratives would have been detectable: references to security incidents increasingly co-occurring with oil supply, shipping insurance, and inflation concerns. Second, emotional indicators would likely have shown a transition from localized fear to systemic anxiety, reflected in broader economic discourse. Third, endogeneity measures would have revealed that attention dynamics were becoming increasingly self-sustained, with narratives propagating across financial, geopolitical, and industrial domains even in the absence of new triggering events.

Once the Strait was effectively disrupted, the response revealed the system's criticality. Oil prices surged dramatically, supply chains fractured, and inflationary pressures spread globally . The amplitude and persistence of the response, far exceeding that of previous oil shocks, demonstrated that the system had entered a critical or possibly even super-critical regime. Even temporary reopenings failed to restore stability, as underlying structural vulnerabilities remains.

From an analytical perspective, the crisis illustrates a key mechanism: the transformation of a localized geopolitical event into a **global dragon-king** through endogenous amplification across tightly coupled systems, such as energy, key commodities (fertilizers, Helium), finance, trade, and political narratives. The observatory's role is precisely to identify when such coupling intensifies to the point that a regional perturbation can propagate globally.

The Chinese use of the Iran conflict as a real-time data laboratory reinforces this interpretation. While the crisis unfolded, AI systems are simultaneously extracting structured signals from the same environment to anticipate future actions. This duality, namely crisis as event and crisis as data-generating process, highlights the fundamental shift: geopolitical systems are no longer only experienced; they are continuously measured, modeled, and anticipated. A Geopolitical Crisis Observatory should therefore be understood as institutionalizing this second layer within the media sphere.

## 11. Conclusion

The central contribution that complex-systems science can offer media institutions for geopolitical foresight is not a new rhetoric of prediction but a disciplined way of thinking about susceptibility, amplification, and regime change. Critical events should be defined as transitions produced by the interaction of endogenous maturation and exogenous perturbation. Weak signals should be identified not by volume alone but by dynamical structure. Endogeneity should be measured, not merely invoked. Dragon-kings should be recognized as regime-generated extremes rather than mysterious surprises. Non-normal dynamics should be incorporated because social and political systems may collapse along transiently amplified directions before classical instability becomes obvious. And organizational concealment should be treated as part of the causal system, not as an afterthought.

The emergence of AI-driven observatories in strategic contexts demonstrates that this transformation is already underway outside the media domain. The question is no longer whether such observatories are possible, but whether media organizations will develop them in time to remain relevant in an environment where anticipation becomes the primary source of informational advantage. The Geopolitical Crisis Observatory thus represents not only a scientific framework, but a necessary evolution of journalism in the age of AI and systemic risk.